\documentclass[prd,twocolumn,showpacs,superscriptaddress,preprintnumbers,%
floatfix,amsmath,amssymb,amsfonts]{revtex4}

\usepackage{graphicx}
\usepackage{dcolumn}
\usepackage{bm}
\usepackage{subfigure}

\RequirePackage{xspace}
\usepackage{relsize}

\def\babar{\mbox{\slshape B\kern-0.1em{\smaller A}\kern-0.1em
    B\kern-0.1em{\smaller A\kern-0.2em R}}}
\def\CP                {\ensuremath{C\!P}\xspace}
\def\to                 {\ensuremath{\rightarrow}\xspace}

\def\ccbar {\ensuremath{c\overline c}\xspace}

\def\piz   {\ensuremath{\pi^0}\xspace}

\def\pip   {\ensuremath{\pi^+}\xspace}
\def\pim   {\ensuremath{\pi^-}\xspace}

\def\Kbar  {\kern 0.2em\overline{\kern -0.2em K}{}\xspace}

\def\Kz    {\ensuremath{K^0}\xspace}
\def\Kzb   {\ensuremath{\Kbar^0}\xspace}
\def\KzKzb {\ensuremath{\Kz \kern -0.16em \Kzb}\xspace}
\def\Kp    {\ensuremath{K^+}\xspace}
\def\Km    {\ensuremath{K^-}\xspace}

\def\KpKm  {\ensuremath{\Kp \kern -0.16em \Km}\xspace}
\def\KS    {\ensuremath{K^0_{\scriptscriptstyle S}}\xspace}

\def\Dbar    {\kern 0.2em\overline{\kern -0.2em D}{}\xspace}

\def\Dz      {\ensuremath{D^0}\xspace}
\def\Dzb     {\ensuremath{\Dbar^0}\xspace}
\def\Dp      {\ensuremath{D^+}\xspace}
\def\Dm      {\ensuremath{D^-}\xspace}

\def\Dmp     {\ensuremath{D^\mp}\xspace}
\def\DpDm    {\ensuremath{\Dp {\kern -0.16em \Dm}}\xspace}
\def\Dstar   {\ensuremath{D^*}\xspace}

\def\Dstarzb {\ensuremath{\Dbar^{*0}}\xspace}
\def\Dstarp  {\ensuremath{D^{*+}}\xspace}
\def\Dstarm  {\ensuremath{D^{*-}}\xspace}
\def\Dstarpm {\ensuremath{D^{*\pm}}\xspace}

\def\Dstapm  {\ensuremath{D^{(*)\pm}}\xspace}

\def\Bbar    {\kern 0.18em\overline{\kern -0.18em B}{}\xspace}

\def\BB      {\ensuremath{B\Bbar}\xspace} 
\def\Bz      {\ensuremath{B^0}\xspace}
\def\Bzb     {\ensuremath{\Bbar^0}\xspace}
\def\BzBzb   {\ensuremath{\Bz {\kern -0.16em \Bzb}}\xspace}
\def\Bu      {\ensuremath{B^+}\xspace}

\def\Bp      {\ensuremath{\Bu}\xspace}

\mathchardef\Upsilon="7107
\def\Y#1S{\ensuremath{\Upsilon{(#1S)}}\xspace}
\def\FourS {\Y4S}

\def\mes        {\mbox{$m_{\rm ES}$}\xspace}

\newcommand{\tev}{\ensuremath{\mathrm{\,Te\kern -0.1em V}}\xspace}
\newcommand{\gev}{\ensuremath{\mathrm{\,Ge\kern -0.1em V}}\xspace}
\newcommand{\mev}{\ensuremath{\mathrm{\,Me\kern -0.1em V}}\xspace}
\newcommand{\kev}{\ensuremath{\mathrm{\,ke\kern -0.1em V}}\xspace}
\newcommand{\ev}{\ensuremath{\mathrm{\,e\kern -0.1em V}}\xspace}
\newcommand{\gevc}{\ensuremath{{\mathrm{\,Ge\kern -0.1em V\!/}c}}\xspace}
\newcommand{\mevc}{\ensuremath{{\mathrm{\,Me\kern -0.1em V\!/}c}}\xspace}
\newcommand{\gevcc}{\ensuremath{{\mathrm{\,Ge\kern -0.1em V\!/}c^2}}\xspace}
\newcommand{\mevcc}{\ensuremath{{\mathrm{\,Me\kern -0.1em V\!/}c^2}}\xspace}
\def\ps   {\ensuremath{\rm \,ps}\xspace}

\def\pep2{PEP-II}
\def\BF{$B$ Factory}

\newcommand{\stat}{\ensuremath{\mathrm{(stat)}}\xspace}
\newcommand{\syst}{\ensuremath{\mathrm{(syst)}}\xspace}

\def\stwob{\ensuremath{\sin\! 2 \beta   }\xspace}
\def\mistag{\ensuremath{w}\xspace}
\def\deltaz{\ensuremath{{\rm \Delta}z}\xspace}
\def\deltat{\ensuremath{{\rm \Delta}t}\xspace}
\def\deltamd{\ensuremath{{\rm \Delta}m_d}\xspace}

\def\Bztodd     {\ensuremath{\Bz \to \DpDm}\xspace}
\def\Bztodstd   {\ensuremath{\Bz \to \Dstarp \Dm}\xspace}
\def\Bztodstdst {\ensuremath{\Bz \to \Dstarp \Dstarm}\xspace}
\def\Bztoddst   {\ensuremath{\Bz \to \Dp \Dstarm}\xspace}
\def\Bztodstadsta   {\ensuremath{\Bz \to D^{(*)+} D^{(*)-}}\xspace}
\def\Rt  {\ensuremath{R_{\perp}}\xspace}
\def\massLik {\ensuremath{\mathcal{L}_\text{mass}}\xspace}

\def\thtr {\ensuremath{\theta_\text{tr}}\xspace}
\def\costhtr {\ensuremath{\cos\thtr}\xspace}

\begin{document}

\preprint{{\babar}-PUB-08/039}
\preprint{SLAC-PUB-13327}
\preprint{arXiv:0808:1866 [hep-ex]}

\title{\boldmath Measurements of time-dependent \CP asymmetries in \Bztodstadsta decays}

\author{B.~Aubert}
\author{M.~Bona}
\author{Y.~Karyotakis}
\author{J.~P.~Lees}
\author{V.~Poireau}
\author{E.~Prencipe}
\author{X.~Prudent}
\author{V.~Tisserand}
\affiliation{Laboratoire de Physique des Particules, IN2P3/CNRS et Universit\'e de Savoie, F-74941 Annecy-Le-Vieux, France }
\author{J.~Garra~Tico}
\author{E.~Grauges}
\affiliation{Universitat de Barcelona, Facultat de Fisica, Departament ECM, E-08028 Barcelona, Spain }
% ...
\collaboration{The \babar\ Collaboration}
\noaffiliation

\date{\today}

\begin{abstract}
  We present new measurements of time-dependent \CP asymmetries for
  \Bztodstadsta decays using $(467\pm 5)\times 10^6$ \BB pairs
  collected with the \babar\ detector located at the \pep2\ \BF\ at
  the Stanford Linear Accelerator Center.  We determine the \CP-odd
  fraction of the \Bztodstdst decays to be $\Rt = 0.158 \pm 0.028 \pm
  0.006$ and find \CP asymmetry parameters $S_+ = -0.76 \pm 0.16 \pm
  0.04$ and $C_+ = +0.00 \pm 0.12 \pm 0.02$ for the \CP-even component
  of this decay and $S_\perp = -1.80 \pm 0.70 \pm 0.16$ and $C_\perp =
  +0.41 \pm 0.49 \pm 0.08$ for the \CP-odd component.  We measure $S =
  -0.63 \pm 0.36 \pm 0.05$ and $C = -0.07 \pm 0.23 \pm0.03$ for
  \Bztodd, $S = -0.62 \pm 0.21 \pm 0.03$ and $C = +0.08 \pm 0.17 \pm
  0.04$ for \Bztodstd, and $S = -0.73 \pm 0.23 \pm 0.05$ and $C =
  +0.00 \pm 0.17 \pm 0.03$ for \Bztoddst.  For the $\Bz\to
  D^{*\pm}D^{\mp}$ decays, we also determine the \CP-violating
  asymmetry $\mathcal{A} = +0.008 \pm 0.048 \pm 0.013$.  In each case,
  the first uncertainty is statistical and the second is systematic.
  The measured values for the asymmetries are all consistent with the
  Standard Model.
\end{abstract}

\maketitle

\section{Introduction}
\label{sec:intro}

In the Standard Model (SM), \CP violation is described by the
Ca\-bib\-bo-Ko\-ba\-ya\-shi-Mas\-ka\-wa (CKM) quark mixing matrix,
$V$~\cite{Cabibbo:1963yz,Kobayashi:1973fv}.  In particular, an
irreducible complex phase in the $3\times3$ mixing matrix is the
source of all SM \CP violation. Both the
\babar~\cite{PhysRevLett.87.091801} and Belle
\cite{PhysRevLett.87.091802} collaborations have measured the \CP
parameter \stwob, where $\beta\equiv
\text{arg}\left[-V_{cd}V_{cb}^*/V_{td}V_{tb}^*\right]$, in $b\to
(\ccbar)s$ processes.

\begin{figure}[bt]
\subfigure[\label{fig:tree}~Tree]{\includegraphics[width=0.475\columnwidth]{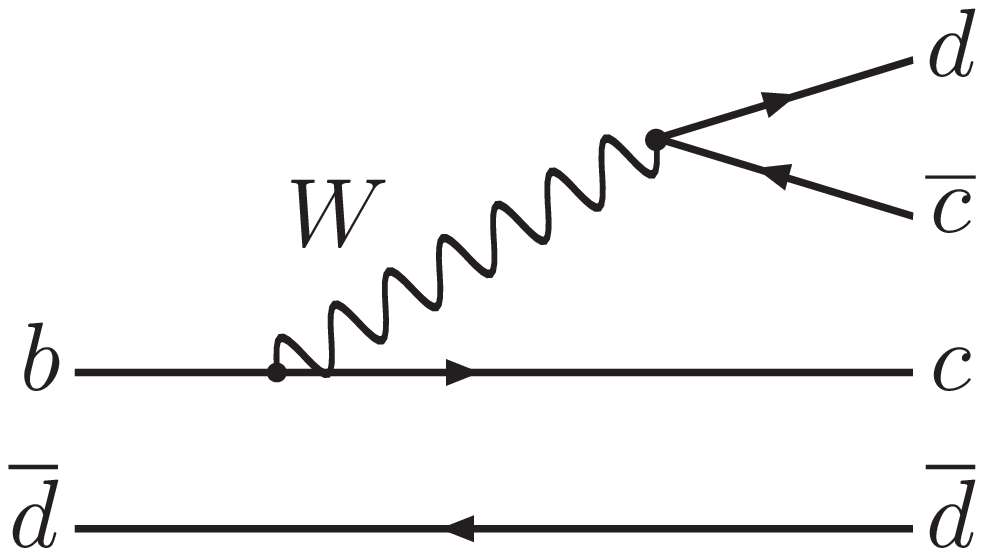}}\quad
\subfigure[\label{fig:penguin}~Penguin]{\includegraphics[width=0.475\columnwidth]{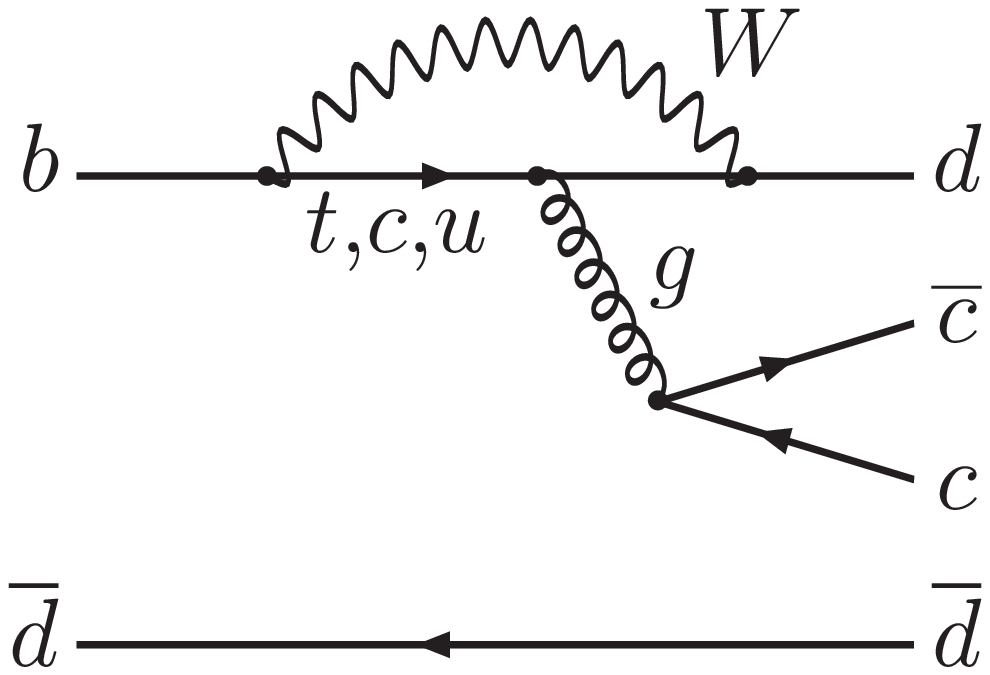}}
\caption{\label{fig:diags}Leading-order Feynman graphs for the $\Bzb\to D^{(*)+} D^{(*)-}$ decays.}
\end{figure}

The leading-order diagrams contributing to \Bztodstadsta decays are
shown in Fig.~\ref{fig:diags}, where the color-favored tree-diagram of
Fig~\ref{fig:tree} dominates.  When neglecting the penguin (loop)
amplitude in Fig.~\ref{fig:penguin}, the mixing-induced \CP asymmetry
of \Bztodstadsta, denoted $S$, is also determined by
\stwob~\cite{Sanda:1996pm}.  The effect of neglecting the penguin
amplitude has been estimated in models based on factorization and
heavy quark symmetry, and the corrections are expected to be a few
percent \cite{Xing:1998ca,Xing:1999yx}.  Large deviations of $S$ in
\Bztodstadsta decays with respect to \stwob determined from
$b\to(\ccbar)s$ transitions could indicate physics beyond the
SM~\cite{Grossman:1996ke,Gronau:2008ed,Zwicky:2007vv}.

The \CP asymmetries of \Bztodstadsta decays have been studied by both
the \babar~\cite{Aubert:2007pa,Aubert:2007rr} and
Belle~\cite{Miyake:2005qb,Aushev:2004uc,Fratina:2007zk}
collaborations.  In the SM, the direct \CP asymmetry $C$, defined in
Sec.~\ref{sec:tdcp}, for the \Bztodstadsta decays is expected to be
near zero.  The Belle Collaboration has observed a 3.2 sigma deviation
of $C$ from zero in the \Bztodd channel~\cite{Fratina:2007zk}.  This
has not been observed by \babar\ nor has it been seen in other
\Bztodstadsta decay modes, which involve the same quark-level
diagrams.  As was pointed out in~\cite{Gronau:2008ed}, understanding
any possible asymmetries in these decays is important to constraining
theoretical models.

In this article, we update the previous measurements of \CP asymmetry
parameters in \Bztodstadsta decays~\cite{Aubert:2007pa,Aubert:2007rr},
including the \CP-odd fraction for \Bztodstdst, using the final
\babar\ data sample.  Charge conjugate decays are included implicitly
in expressions throughout this article unless otherwise indicated.

\section{Detector, data sample, and reconstruction}

\subsection{The {\babar\ }detector}
\label{sec:det}

The data used in this analysis were collected with the \babar\
detector~\cite{Aubert:2001tu} operating at the \pep2\ \BF\ located at
the Stanford Linear Accelerator Center (SLAC).  The \babar\ dataset
comprises $(467\pm 5)\times 10^6$ \BB pairs collected from 1999 to
2007 at the center-of-mass (CM) energy $\sqrt{s}=10.58 \gev$,
corresponding to the \FourS resonance.  We use
GEANT4-based~\cite{Agostinelli:2002hh} Monte Carlo (MC) simulation to
study backgrounds and to validate the analysis procedures.

The asymmetric energies of the \pep2\ beams provide an ideal
environment to study time-dependent \CP phenomena in the $\Bz-\Bzb$
system by boosting the \FourS in the laboratory frame, thus making
possible precise determination of the decay vertices of the two $B$
meson daughters.  \babar\ employs a five-layer silicon vertex tracker
(SVT) close to the interaction region to provide precise vertex
measurements and to track low momentum charged particles.  A drift
chamber (DCH) provides excellent momentum measurement of charged
particles.  Particle identification of kaons and pions is primarily
derived from ionization losses in the SVT and DCH and from
measurements of photons produced in the fused silica bars of a
ring-imaging Cherenkov light detector (DIRC).  A CsI(Tl) crystal-based
electromagnetic calorimeter enables reconstruction of photons and
identification of electrons.  All of these systems operate within a
1.5 T superconducting solenoid, whose iron flux return is instrumented
to detect muons.

\subsection{Candidate reconstruction and selection}
\label{sec:candyield}

The candidates used in this analysis are formed from oppositely
charged $D^{(*)}$ mesons where we include the $\Dstarp$ decay modes
$\Dstarp\to\Dz\pip$ and $\Dstarp\to\Dp\piz$ and $D$ decay modes
$\Dz\to\Km\pip$, $\Dz\to\Km\pip\piz$, $\Dz\to\Km\pip\pim\pip$,
$\Dz\to\KS\pip\pim$, and $\Dp\to\Km\pip\pip$.  In the \Bztodstdst
mode, we reject \Bz candidates where both \Dstar mesons decay to
$D\piz$ because of its smaller branching fraction and larger
backgrounds.  Reference~\cite{Aubert:2006ia} contains the details of
the reconstruction procedure, outlined here, used to select signal
candidates.  Charged kaon candidates must be identified as such using
a likelihood technique based on the opening angle of the Cherenkov
light measured in the DIRC and the ionization energy loss measured in
the SVT and DCH~\cite{Aubert:2001tu}.  We reconstruct \KS candidates
from two oppositely charged tracks, geometrically constrained to a
common vertex and with an invariant mass within $20 \mev$ of the
nominal value~\cite{Yao:2006px}.  We also require that the $\chi^2$
probability of the vertex fit of the \KS be greater than 0.1\%.  We
form \piz candidates from a pair of photons detected in the
calorimeter, each with energy greater than $40 \mev$.  The invariant
mass of the two photons must be less than $30 \mevcc$ from the nominal
\piz mass, and their summed energy must be greater than $200 \mev$.
In addition, we apply a mass constraint to the \piz candidates.  We
require the reconstructed $D$ meson candidate mass to be within $20
\mevcc$ of the nominal value, except for the $\Dz\to\Km\pip\piz$
decays where we use a looser requirement of $40 \mevcc$.  The
daughters of each $D$ candidate are fit to a common vertex with their
combined mass constrained to that of the $D$ meson.  We use $D$
candidates combined with a pion track with momentum less than $450
\mevc$ in the CM frame to form \Dstarp candidates.  We fit the \Bz
decay with a vertex constraint.

Since the time of our previous
publications~\cite{Aubert:2007pa,Aubert:2007rr,Aubert:2006ia}, the
\babar\ reconstruction routines have been extensively revised, leading
to significant improvements in localizing and reconstructing tracks,
particularly for low momentum charged particles.  These improvements
have increased the reconstruction efficiency for final states with
multiple slow particles, such as the \Bztodstdst channel which has a
better than 20\% improvement.  As a result, the statistical
sensitivity of the measurements in this paper has increased more than
would be expected by just the increment in luminosity.

To suppress $e^+e^- \to q\overline{q}$ ($q=u,d,s,\text{ and } c$)
continuum background, we exploit the spherical shape of \BB events by
requiring the ratio of second to zeroth order Fox-Wolfram
moments~\cite{Fox:1978vu} to be less than $0.6$.  We select the \Bz
candidates based on four variables: $\Delta E \equiv E^*_{B} -
\sqrt{s}/2$, where $E^*_{B}$ is the energy of the $B$ meson in the CM
frame, the $D$ candidate flight length significance, defined as the
sum of the two $D$ candidate flight lengths divided by the error on
the sum, a Fisher discriminant~\cite{Asner:1995hc}, and a mass
likelihood of the $D^{(*)}$ mesons.  The Fisher discriminant is a
linear combination of 11 variables: the momentum flow in nine
concentric cones around the thrust axis of the \Bz candidate, the
angle between the thrust axis and the beam axis, and the angle between
the line-of-flight of the \Bz candidate and the beam axis.  The mass
likelihood is formed from Gaussian functions,
\begin{align}
  \massLik &= G(m_D;m_{D_\text{ PDG}},\sigma_{m_D})
  \times G(m_{\Dbar};m_{\Dbar_\text{PDG}},\sigma_{m_{\Dbar}}) \nonumber \\
  & \times \left[ f_\text{core} G(\Delta m_{\Dstarp};\Delta m_{\Dstarp_\text{PDG}},\sigma_{\Delta m_\text{core}}) \right. \nonumber \\
  & + \left. (1 - f_\text{core}) G(\Delta m_{\Dstarp};\Delta m_{\Dstarp_\text{PDG}},\sigma_{\Delta m_\text{tail}})  \right] \nonumber \\
  & \times \left[ f_\text{core} G(\Delta m_{\Dstarm};\Delta m_{\Dstarm_\text{PDG}},\sigma_{\Delta m_\text{core}}) \right. \nonumber \\
  & + \left. (1 - f_\text{core}) G(\Delta m_{\Dstarm};\Delta
    m_{\Dstarm_\text{PDG}},\sigma_{\Delta m_\text{tail}}) \right]\,,
\end{align}
where the PDG subscript refers to the nominal
value~\cite{Eidelman:2004wy}.  The reconstructed masses and
uncertainties $\sigma_{m_{\Dbar}}$ for the $D$ mesons prior to the
mass constraint are used in the likelihood.  The \Dstar portion of the
likelihood is the sum of two Gaussian functions, a central core and a
wider tail.  The value of $f_\text{core}$ and the widths of the \Dstar
Gaussian functions are taken from detailed signal MC studies, which
show good agreement between data and MC samples.  The selection
criteria are optimized for each $D$ decay channel to maximize the
total signal significance $S/\sqrt{S+B}$ for each \Bz decay mode,
where $S$ and $B$ are the signal and background yields, respectively.
The optimized selections are specified in \cite{Aubert:2006ia}.  We
keep candidates with $\mes \equiv \sqrt{s/4-p^{*2}_B} > 5.23 \gevcc$,
where $p^*_B$ is the momentum of the $B$ candidate in the CM frame.
On average 1.1--1.8 candidates per event satisfy all of the selection
criteria depending on the process.  When more than one \Bz candidate
meets the selection criteria, the one with the best \massLik is kept.
We find from MC that this procedure retains the correct candidate more
than 95\% of the time.

\begin{figure}[bt]
\subfigure[\label{fig:mesdstdst}~\Bztodstdst]{\includegraphics[width=0.475\columnwidth]{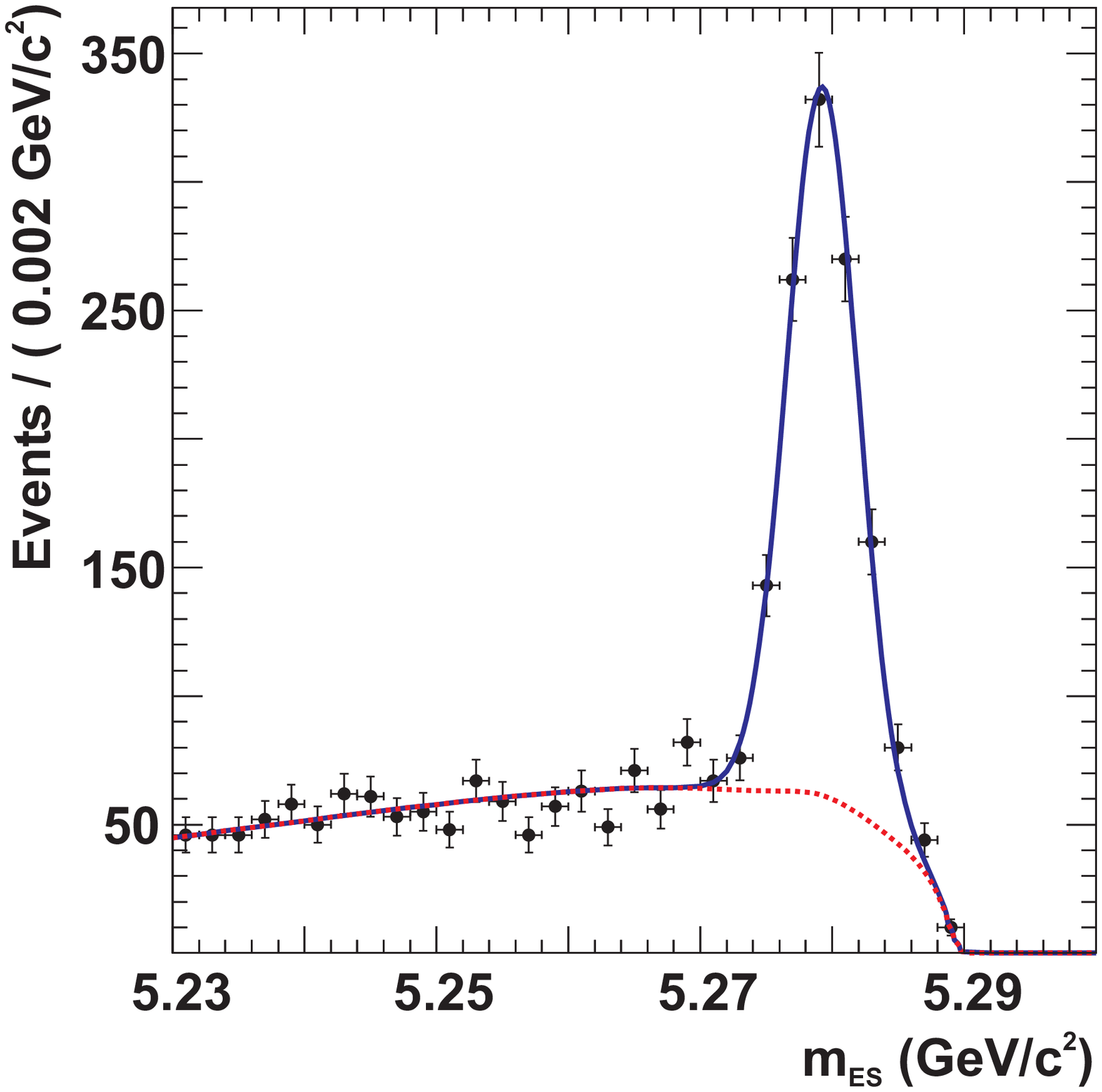}}
\subfigure[\label{fig:mesdpdm}~\Bztodd]{\includegraphics[width=0.475\columnwidth]{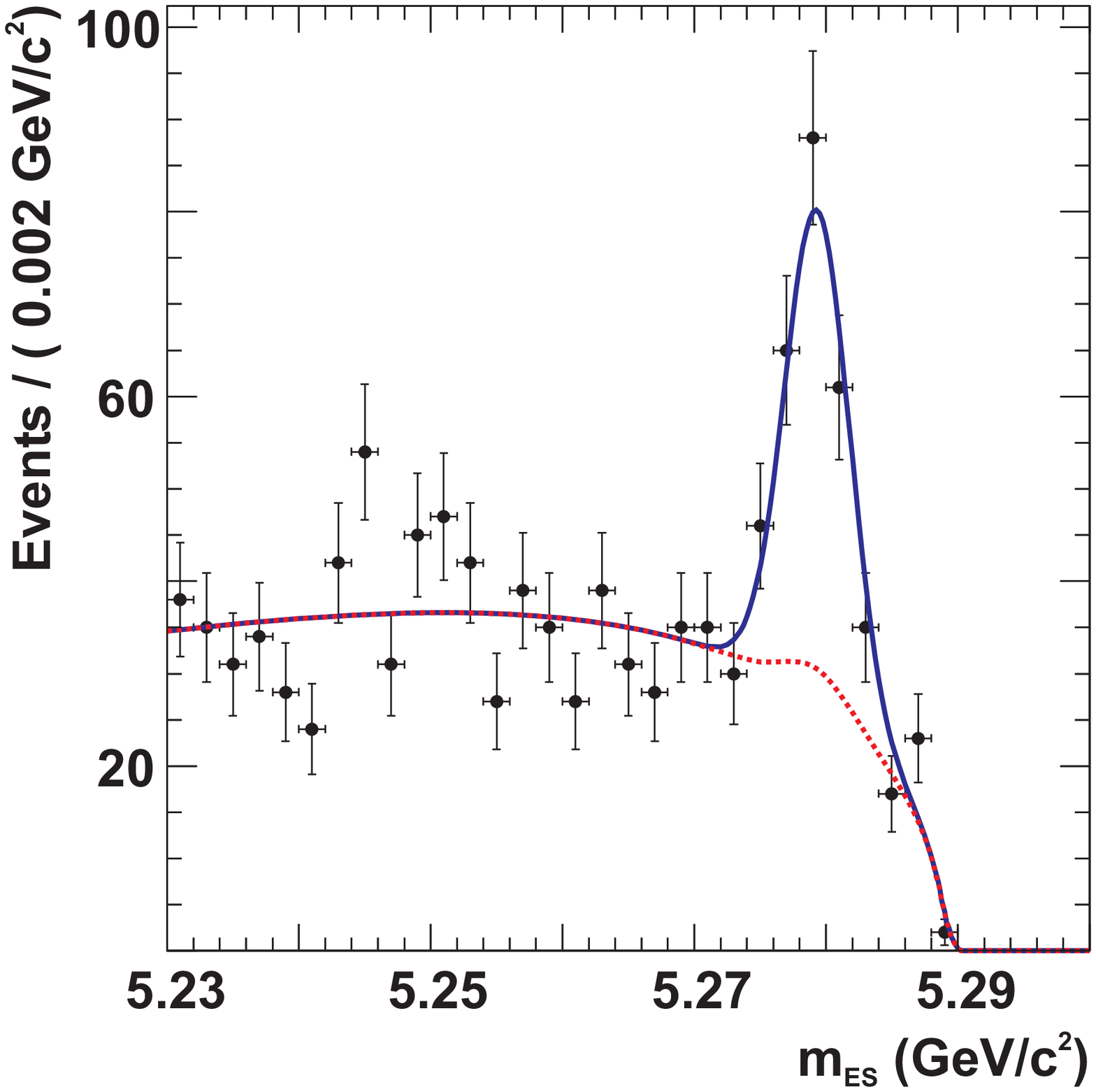}}
\subfigure[\label{fig:mesdstpdm}~\Bztodstd]{\includegraphics[width=0.475\columnwidth]{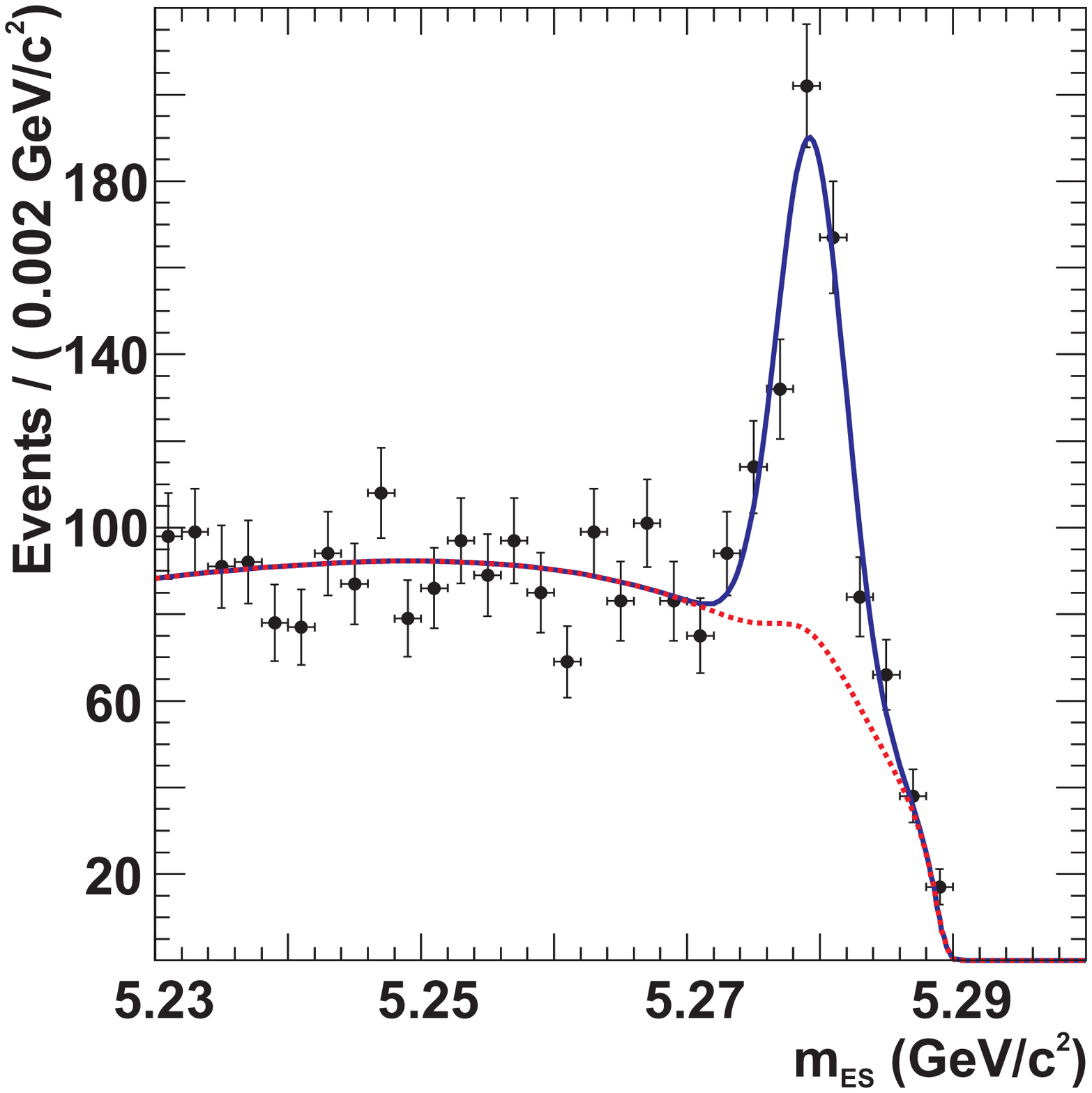}}
\subfigure[\label{fig:mesdstmdp}~\Bztoddst]{\includegraphics[width=0.475\columnwidth]{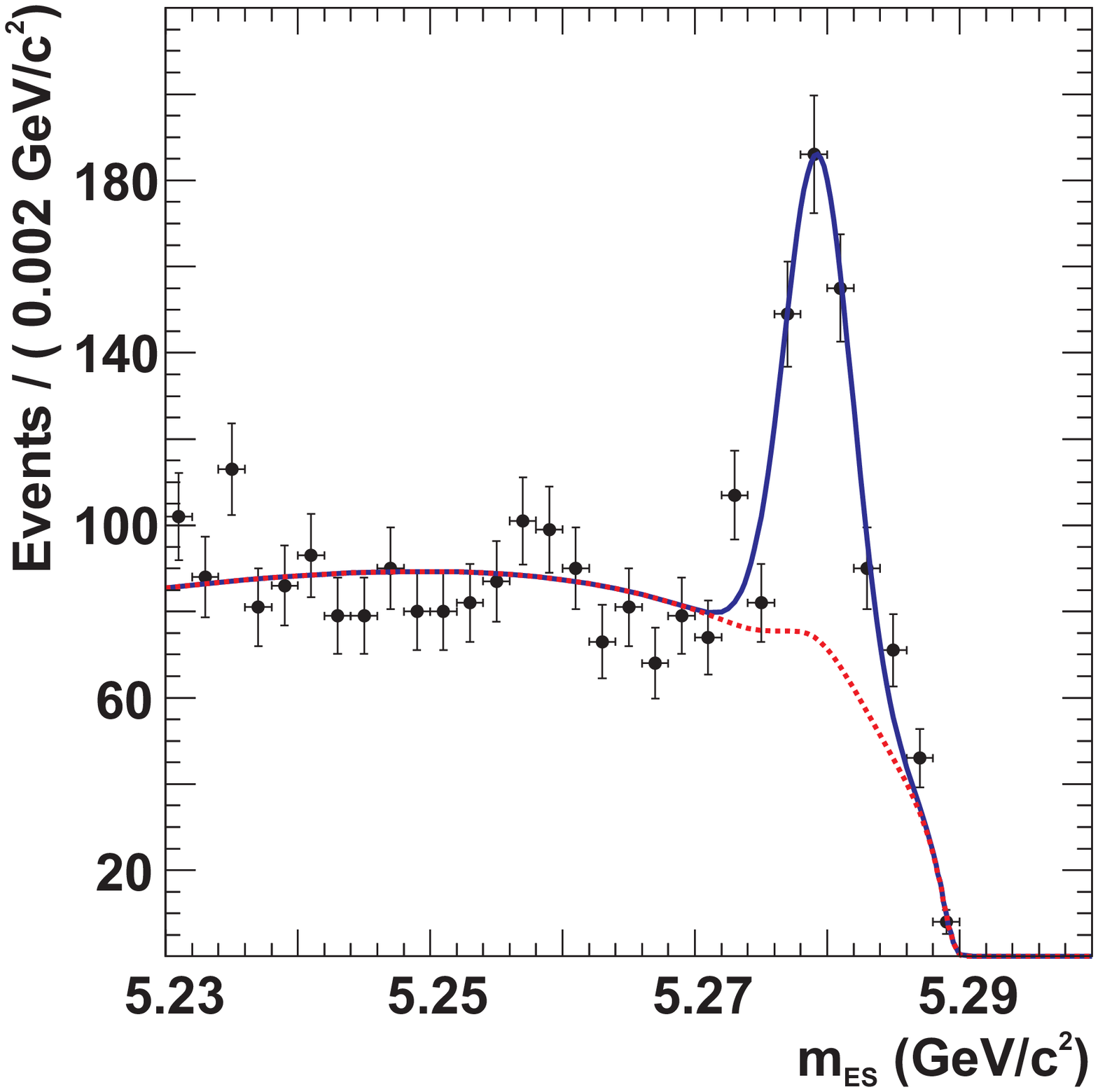}}
\caption{\label{fig:mes}Projections of the \mes fit results.  The
  solid line represents the total fit PDF and the dashed line is the
  background contribution.}
\end{figure}

To determine the signal yields of the data sample, we use unbinned
maximum likelihood (ML) fits to the \mes distributions.  The signal is
described by a Gaussian function and the combinatorial background by a
threshold function~\cite{Albrecht:1990am}.  In detailed MC studies of
the background, we find that there is a background contribution that
exceeds the threshold function in the region $\mes > 5.27 \gevcc$,
where most of the signal events lie.  We describe this component with
a Gaussian function having the same mean and width as the signal and
refer to it as peaking background because if neglected, it would lead
to an overestimate of the signal yields.  In the \Bztodstdst channel,
the peaking background arises primarily from misreconstructed
$\Bp\to\Dstarp\Dstarzb$ events where the slow \piz from the
$\Dstarzb\to\Dzb\piz$ decay is replaced by a \pim to form a \Dstarm
candidate.  For the other three processes, our studies of the
composition of the peaking background show it to be consistent with
that of the combinatorial background in the region $\mes < 5.27
\gevcc$.  We treat the peaking background component as an extension of
the combinatorial background.  The peaking background yields relative
to the signal are fixed from MC to $(1.6\pm 1.9)\%$, $(7.1\pm5.9)\%$,
and $(7.4\pm 2.9)\%$ for the \Bztodstdst, \Bztodd, and
$\Bz\to\Dstarpm\Dmp$ modes, respectively, where the errors are due
primarily to the size of the MC sample available for background
studies.  The signal mean and background shape are free parameters in
the fits.  We fix the width of the signal Gaussian shape for \Bztodd
and $\Bz\to\Dstarpm\Dmp$ to 2.46 \mevcc and 2.55 \mevcc, respectively,
determined from MC, while the width of the \Bztodstdst signal is
allowed to float because of its much higher purity.  The signal yields
are $934 \pm 40$ \Bztodstdst events, $152 \pm 17$ \Bztodd events, $365
\pm 26$ \Bztodstd events, and $359 \pm 26$ \Bztoddst events, where the
uncertainty is statistical only.  The signal yields are consistent
with previously measured \Bztodstadsta decay branching fractions from
\babar~\cite{Aubert:2006ia} and
Belle~\cite{Fratina:2007zk,Abe:2002zy}.  When compared with past
\babar\ measurements~\cite{Aubert:2006ia,Aubert:2005hs,Aubert:2003ca},
the low \Bztoddst yield in Ref.~\cite{Aubert:2007pa} is consistent
with a statistical fluctuation.  The fit projections for each mode
onto \mes are shown in Fig.~\ref{fig:mes}.

\section{\boldmath Time-integrated measurement of the \CP-odd fraction}
\label{sec:rt}

\begin{figure}[tb]
\includegraphics[width=0.95\columnwidth]{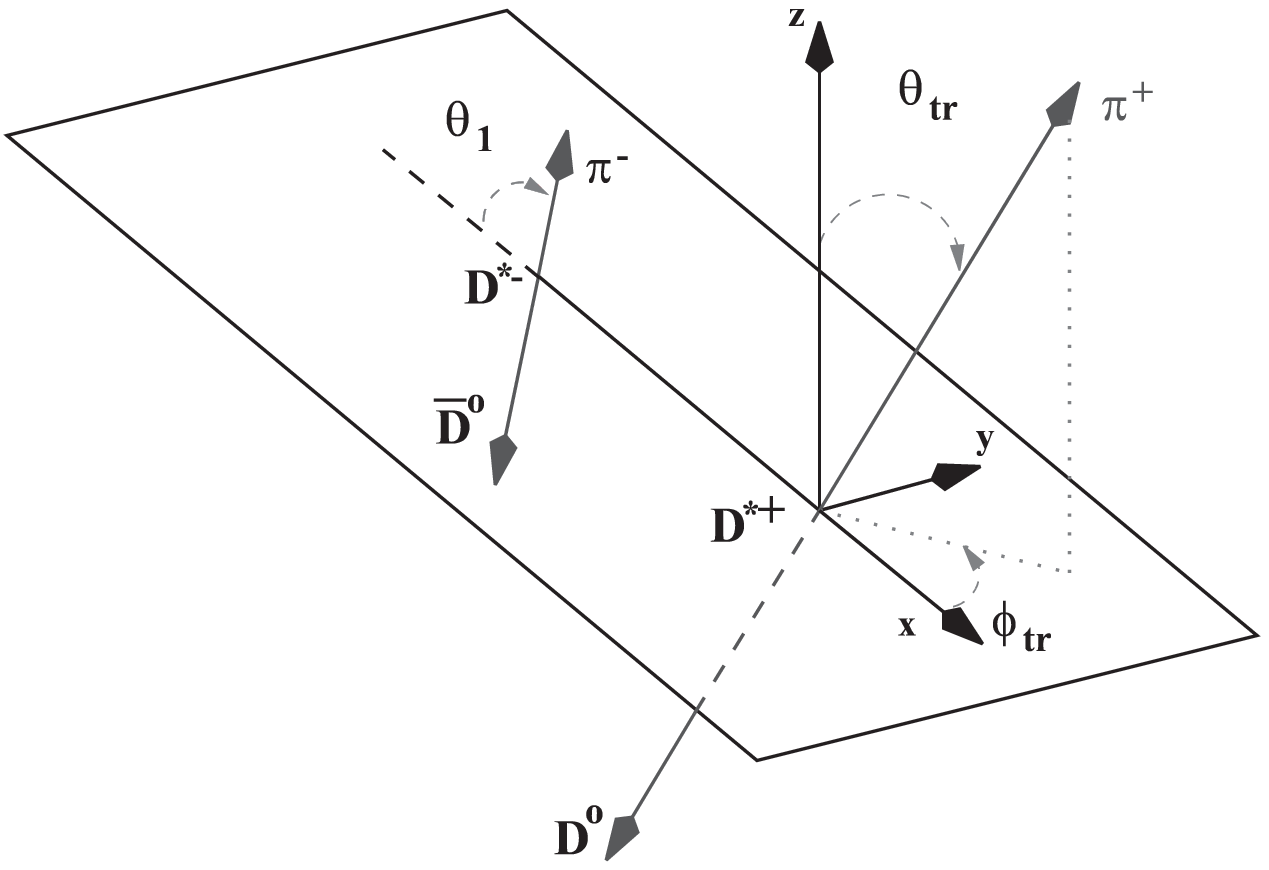}
\caption{\label{fig:trans}Depiction of the \Bztodstdst decay in the
transversity basis with the $\Dstarm\to\Dzb\pim$ decay plane shown.
The three transversity angles are defined in the text.}
\end{figure}

The \Bztodstdst process has two vector mesons in the final state and
is an admixture of \CP-even and \CP-odd states depending on the
orbital angular momentum of the decay products.  We measure the
\CP-odd fraction \Rt using a time-integrated angular
analysis~\cite{Dunietz:1990cj}.  We define the three angles in the
transversity basis as depicted in Fig.~\ref{fig:trans}: the angle
$\theta_1$ between the slow pion from the \Dstarm and the direction
opposite to the \Dstarp momentum in the \Dstarm rest frame; the polar
angle $\thtr$ and the azimuthal angle $\phi_\text{tr}$ of the slow
pion from the \Dstarp in the \Dstarp rest frame where the $z$ axis is
normal to the \Dstarm decay plane and the $x$ axis is opposite the
\Dstarm momentum.  Working in the transversity basis, the
time-dependent angular distribution of the \Bz decay products is
\begin{align}
  \frac{1}{\Gamma}&\frac{{\rm d}^4\Gamma}{{\rm d}\cos\theta_1{\rm
      d}\costhtr{\rm d}\phi_\text{tr}\text{d}t} =
  \frac{9}{16\pi}\frac{1}{\lvert A_0\rvert^2+\lvert
    A_{\|}\rvert^2+\lvert A_{\perp}\rvert^2}
  \times \nonumber \\
  & \left\{ 2\cos^2\theta_1\sin^2\thtr\cos^2\phi_\text{tr}\lvert A_0\rvert^2 \phantom{\frac{1}{2}}\right. \nonumber \\
  &+\sin^2\theta_1\sin^2\thtr\sin^2\phi_\text{tr}\lvert A_{\|}\rvert^2 \nonumber \\
  &+\sin^2\theta_1\cos^2\thtr\lvert A_{\perp}\rvert^2 \nonumber \\
  &-\sin^2\theta_1\sin 2\thtr\sin\phi_\text{tr}\,\text{Im}(A_{\|}^*A_{\perp}) \nonumber \\
  &+\frac{1}{\sqrt{2}}\sin 2\theta_1\sin^2\thtr\sin 2\phi_\text{tr}\,\text{Re}(A_0^*A_{\|}) \nonumber \\
  & \left. -\frac{1}{\sqrt{2}}\sin 2\theta_1\sin
    2\thtr\cos\phi_\text{tr}\,\text{Im}(A_0^*A_{\perp}) \right\}~,
\label{eq:transdiffdecayrate}
\end{align}
where $A_k$, with $k=\|,0,\perp$, represent time-dependent amplitudes
given by
\begin{align}
  A_{k} (t) &= \frac{\sqrt{2}A_{k}(0)}{1+|\lambda_{k}|^2}
  e^{-imt} e^{-t/2\tau_{\Bz}} \nonumber \\
  & \quad \times \left(\cos\frac{\deltamd t}{2} + i \eta_{\CP}^{k}
    \lambda_{k}\sin\frac{\deltamd t}{2}\right)\,.
\label{eq:As}
\end{align}
Here, $\eta_{\CP}^k$ is the \CP eigenvalue, $+1$ for
$A_{\parallel,0}$, $-1$ for $A_\perp$; $\lambda_k$ is the \CP
parameter defined in Sec.~\ref{sec:tdcp}; \deltamd is the \Bz mixing
frequency, $(0.507\pm0.005) \ps^{-1}$; and $\tau_{\Bz}$ is the \Bz
lifetime, $(1.530\pm0.009) \ps$~\cite{Yao:2006px}.  Expressions
similar to Eq.~\ref{eq:transdiffdecayrate} hold for \Bzb decays where
each $A_k$ is replaced by the appropriate $\overline{A}_k$ including
$A_\perp \to -\overline{A}_\perp$.  Integrating
Eq.~\ref{eq:transdiffdecayrate} over $t$, $\phi_\text{tr}$,
$\cos\theta_1$ and averaging over $B$ flavor while taking into account
detector efficiency yields
\begin{align}
{\frac{1}{\Gamma}} \frac{{\rm d}\Gamma}{{\rm d}\costhtr} &= 
{\frac{9}{32\pi}} (1-\Rt) \sin^2 \thtr \nonumber \\
&\times \left\{ 
\frac{1+\alpha}{2} I_{0}(\costhtr) + 
\frac{1-\alpha}{2} I_{\parallel}(\costhtr) 
\right\} \nonumber \\
&+ {\frac{3}{2}} \Rt \cos^2 \thtr \times I_{\perp}(\costhtr)\,,
\label{eq:angdist}
\end{align}
where we define
\begin{align*}
  \Rt & = \frac{|A^0_{\perp}|^2}{|A^0_0|^2 + |A^0_{\parallel}|^2 + |A^0_{\perp}|^2} \\
  \alpha & =  \frac{|A^0_{0}|^2  - |A^0_{\parallel}|^2 }{|A^0_0|^2 + |A^0_{\parallel}|^2}\,,
\end{align*}
and $A_{k}^0 = A_k(0)$.  The three efficiency moments $I_k(\costhtr)$
are defined as
\begin{equation}
  I_{k}(\costhtr) =  \int \mathrm{d}\cos\theta_1 \mathrm{d}\phi_\text{tr}
  g_k(\theta_1,\phi_\text{tr})\varepsilon(\theta_1,\thtr,\phi_\text{tr})\,,
  \label{eq:moments}
\end{equation}
where $g_0 = 4\cos^2\theta_1 \cos^2\phi_\text{tr}$,
$g_\parallel=2\sin^2\theta_1\sin^2\phi_\text{tr}$,
$g_\perp=\sin^2\theta_1$, and $\varepsilon$ is the detector
efficiency.  The moments $I_k$ are parameterized as second-order even
polynomials in \costhtr whose parameters are determined from signal MC
simulation and fixed in the fit.  The three $I_k$ functions deviate
only slightly from the same constant, making Eq.~\ref{eq:angdist}
nearly insensitive to $\alpha$, which we fix to zero in the fit.

Because \costhtr is defined with respect to the slow pion from the
\Dstarp decay, the measurement resolution smears its distribution.  We
convolve the function from Eq.~\ref{eq:angdist} with a resolution
function $\mathcal{R}(\Delta\thtr)$ which is modeled as the sum of
three Gaussian functions.  In addition, we include an uncorrelated
Gaussian shape centered at $\pi/2$ and normalized in $0 < \thtr < \pi$
to describe decays where the slow pion is poorly reconstructed leading
to a loss of angular information.  The uncorrelated term represents
3\% of the signal events where both slow pions are charged and around
16\% in the modes where one of the slow pions is neutral.  We
determine the parameters of the resolution model and of the
uncorrelated term from signal MC simulation and fix them in the ML
fit.  Small differences observed in the angular distributions based on
the charge of the slow pions lead us to divide the efficiency moment
and resolution parameters into three categories, $\piz\pim$,
$\pip\piz$, and $\pip\pim$.

\begin{figure}[tb]
\includegraphics[width=0.85\columnwidth]{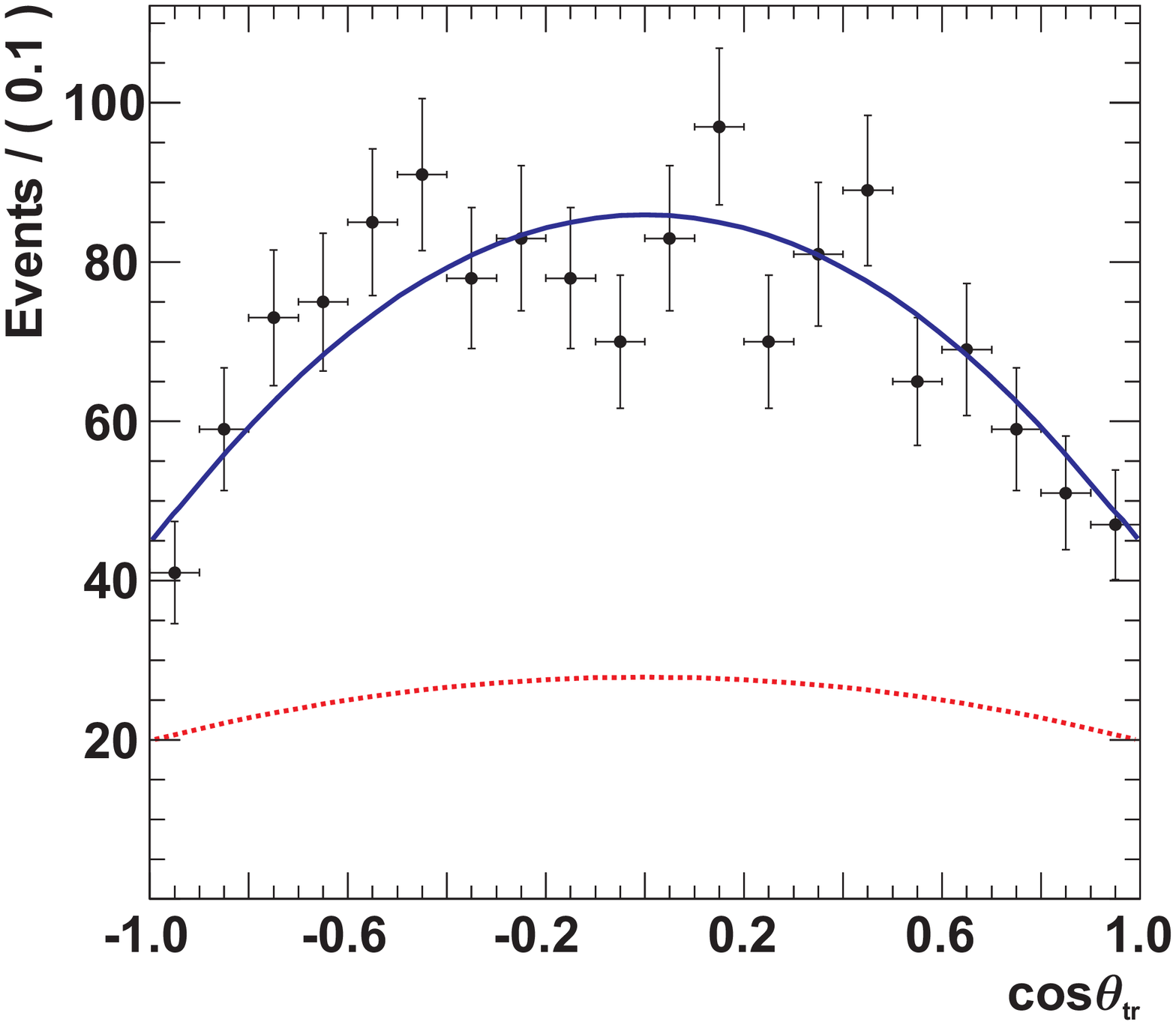}
\caption{\label{fig:rt}Projection of the fit result onto \costhtr
  for events with $\mes>5.27 \gevcc$. The solid line is the projected 
  fit result.  The dashed line is the background component.}
\end{figure}

We determine \Rt in a simultaneous unbinned ML fit to the \mes and
\costhtr distributions for the three slow-pion modes.  The \mes
probability density function (PDF) was described in
Sec.~\ref{sec:candyield}.  The signal \costhtr distribution is given
by Eq.~\ref{eq:angdist} convolved with the resolution model.  The
background \costhtr distribution is modeled as a second-order even
polynomial $f_\text{bg}(\costhtr) = 1 + b_2\cos^2\thtr$, where $b_2$,
common to the three slow-pion modes, is allowed to float.  The yield
for each of the three slow-pion modes is determined by the fit.  We
validate the fitting procedure using high-statistics MC samples
divided into data-sized subsets and find no significant bias.  Fitting
the data and including systematic uncertainties described below, we
find
\begin{equation}
\Rt = 0.158 \pm 0.028\stat \pm 0.006\syst\,.
\end{equation}
Figure~\ref{fig:rt} shows the projection of the fit result.

To evaluate the systematic uncertainty of \Rt, we vary the parameters
used to model the efficiency moments within the uncertainties of the
MC simulation used to extract them.  We do the same for the parameters
used to model the experimental resolution.  In both cases, we take
into account correlations among the parameters when perturbing the
values.  We fix $\alpha$ to zero in the nominal fit, so we also set it
to $\pm 1$ and assign the effect on the fitted result as a systematic
uncertainty.  We change the \mes and \costhtr shapes of the peaking
background and assign the corresponding changes in \Rt as a systematic
uncertainty.  We allow the \costhtr background to have an additional
fourth-order term to test our assumption of this background shape.
This term is found to be consistent with zero, and we take the
difference in \Rt with respect to the nominal second-order background
description as the uncertainty with this model.  We include as a
systematic uncertainty the statistical uncertainty associated with the
MC validation.  A summary of the systematic uncertainties is found in
Table~\ref{tab:rtsyst}.  The total systematic uncertainty is the sum
in quadrature of the individual contributions.

\begin{table}[tb]
\caption{\label{tab:rtsyst}Summary of systematic uncertainties on the 
measurement of \Rt.}
\begin{ruledtabular}
\begin{tabular}{ld}
Angular efficiency moments & 0.0024 \\
Angular measurement resolution & 0.0036 \\
$\alpha$ parameter uncertainty & 0.0026 \\
Peaking background & 0.0014 \\
\costhtr background shape & 0.0002 \\
Potential fit bias & 0.0017 \\
\hline
Total & 0.0055 \\
\end{tabular}
\end{ruledtabular}
\end{table}

\section{\boldmath Time-dependent \CP measurement}
\label{sec:tdcp}

The decay rate $f_+$($f_-$) of the neutral $B$ meson to a common final
state accompanied by a \Bz(\Bzb) tag is
\begin{align}
  f_\pm(\deltat) &\propto e^{-\lvert\deltat\rvert/\tau_{\Bz}}\left\{%
    (1\mp\Delta\mistag)\pm(1-2\mistag)\times \right. \nonumber \\
  & \quad \left.\left[S\sin(\deltamd\deltat) -
      C\cos(\deltamd\deltat)\right]\right\}\,,
  \label{eq:tddecayrate}
\end{align}
with \CP asymmetry parameters $S = 2\text{Im}(\lambda)/(1 +
\lvert\lambda\rvert^2)$, $C = (1 - \lvert\lambda\rvert^2)/(1 +
\lvert\lambda\rvert^2)$, and $\lambda = (q/p) (\overline{A}/A)$, where
$A$ ($\overline{A}$) is the decay amplitude for \Bz (\Bzb) and $q/p$
is the ratio of the flavor contributions to the mass
eigenstates~\cite{Harrison:1998yr}. The parameter \mistag is the
average mistag probability, and $\Delta\mistag$ is the difference
between the mistag probabilities for \Bz and \Bzb.  Here,
$\deltat\equiv t_\text{reco}-t_\text{tag}$ is the proper time
difference between the $B$ reconstructed as \Bztodstadsta
($B_\text{rec}$) and the $B$ used to tag the flavor ($B_\text{tag}$).
In the case of \Bztodstdst, we obtain an expression similar to
Eq.~\ref{eq:tddecayrate} from Eqs.~\ref{eq:transdiffdecayrate}
and~\ref{eq:As},
\begin{align}
  f_\pm(\deltat,\costhtr) &\propto e^{-\lvert\deltat\rvert/\tau_{\Bz}}
  \left\{F(1\mp\Delta\mistag)\pm (1-2\mistag)\times \right. \nonumber \\
  &\left.\left[G\sin(\deltamd\deltat) - H\cos(\deltamd\deltat)\right]
  \right\}\,.
  \label{eq:cprt}
\end{align}
The $F$, $G$, and $H$ coefficients~\cite{CoefDefs} are
\begin{align}
  F &= (1-\Rt)\sin^2\thtr + 2\Rt\cos^2\thtr\,, \nonumber \\
  G &= (1-\Rt)S_+\sin^2\thtr - 2\Rt S_\perp\cos^2\thtr\,, \nonumber \\
  H &= (1-\Rt)C_+\sin^2\thtr + 2\Rt C_\perp\cos^2\thtr\,.
  \label{eq:dtcoefs}
\end{align}
The $\lambda_k$ parameters in Eq.~\ref{eq:As} need not be the same
because of possible differences in the relative contribution of
penguin and tree amplitudes, therefore the $S$ and $C$ parameters for
each of the three $(0,\|, \perp)$ amplitudes can also differ.  Note
that the minus sign before $S_\perp$ in the expression for $G$ absorbs
$\eta_{\CP}^\perp$.  We then define
\begin{equation}
  S_+=\frac{S_\|\lvert A_\|^0\rvert^2 + S_0\lvert A_0^0\rvert^2}%
  {\lvert A_\|^0\rvert^2 + \lvert A_0^0\rvert^2}, \quad%
  C_+=\frac{C_\|\lvert A_\|^0\rvert^2 + C_0\lvert A_0^0\rvert^2}%
  {\lvert A_\|^0\rvert^2 + \lvert A_0^0\rvert^2}\,,
  \label{eq:sc}
\end{equation}
where $A_k^0=A_k(0)$ from Eq.~\ref{eq:As}.

In the absence of penguin contributions, $S_{\Dp\Dm} = S_+ = S_\perp =
-\stwob$, and $C_{\Dp\Dm} = C_+ = C_\perp = 0$.  Because
$\Bz\to\Dstarpm\Dmp$ is not a \CP eigenstate, the expressions for $S$
and $C$ are related, $S_{\Dstarpm\Dmp} = -\sqrt{1 -
  C_{\Dstarpm\Dmp}}\sin(2\beta_\text{eff}\pm\delta)$, where $\delta$
is the strong phase difference between \Bztodstd and \Bztoddst
decays~\cite{Aleksan:1993qk}.  Neglecting the penguin contributions,
$\beta_\text{eff}=\beta$, and $C_{\Dstarp\Dm}=- C_{\Dp\Dstarm}$.

The technique used to measure the time-dependent \CP asymmetry is
discussed in detail in Ref.~\cite{Aubert:2002rg}.  We calculate
\deltat between the two $B$ decays from the measured separation
\deltaz of their decay vertices along the $z$ axis.  The
$B_\text{rec}$ decay vertex is determined from the daughter tracks of
the \Bztodstadsta decay.  The $B_\text{tag}$ decay vertex is
determined in a fit of the charged tracks not belonging to
$B_\text{rec}$ to a common vertex with a constraint on the beamspot
location and the $B_\text{rec}$ momentum.  Events that do not satisfy
$\lvert\deltat\rvert < 20 \ps$ and $\sigma_{\deltat}<2.5 \ps$ are
considered untagged in the time-dependent fit.

The flavor of the $B_\text{tag}$ meson is determined using a
multivariate analysis of its decay products~\cite{Aubert:2002rg}.  The
tagging algorithm classifies the $B$ flavor and assigns the candidate
to one of six mutually exclusive tagging categories based on the
output.  A seventh untagged category is for events where the flavor
could not be determined.  The performance of the tagging algorithm,
its efficiency and mistag rates, is evaluated using the time-dependent
evolution of a high-statistics data sample of $\FourS\to
B_\text{tag}B_\text{flav}$, where the $B_\text{flav}$ meson decays to
a flavor eigenstate $D^{(*)-}h^+$ and $h^+$ may be a \pip, $\rho^+$,
or $a_1^+$.  The tagging algorithm has an efficiency
$\varepsilon_\text{tag} = (74.4\pm 0.1)\%$ and an effective tagging
power $Q \equiv \varepsilon_\text{tag} (1-\mistag)^2=(31.2\pm0.3)\%$.
The finite resolution of the $B$ vertex reconstruction smears the
distributions described in Eqs.~\ref{eq:tddecayrate}
and~\ref{eq:cprt}.  This measurement resolution is modeled as the sum
of three Gaussian functions described in Ref.~\cite{Aubert:2002rg},
the parameters of which are also determined from the $B_\text{flav}$
sample.

\begin{figure*}[bt]
  \subfigure[~\Bztodstdst]{\includegraphics[width=0.75\columnwidth]{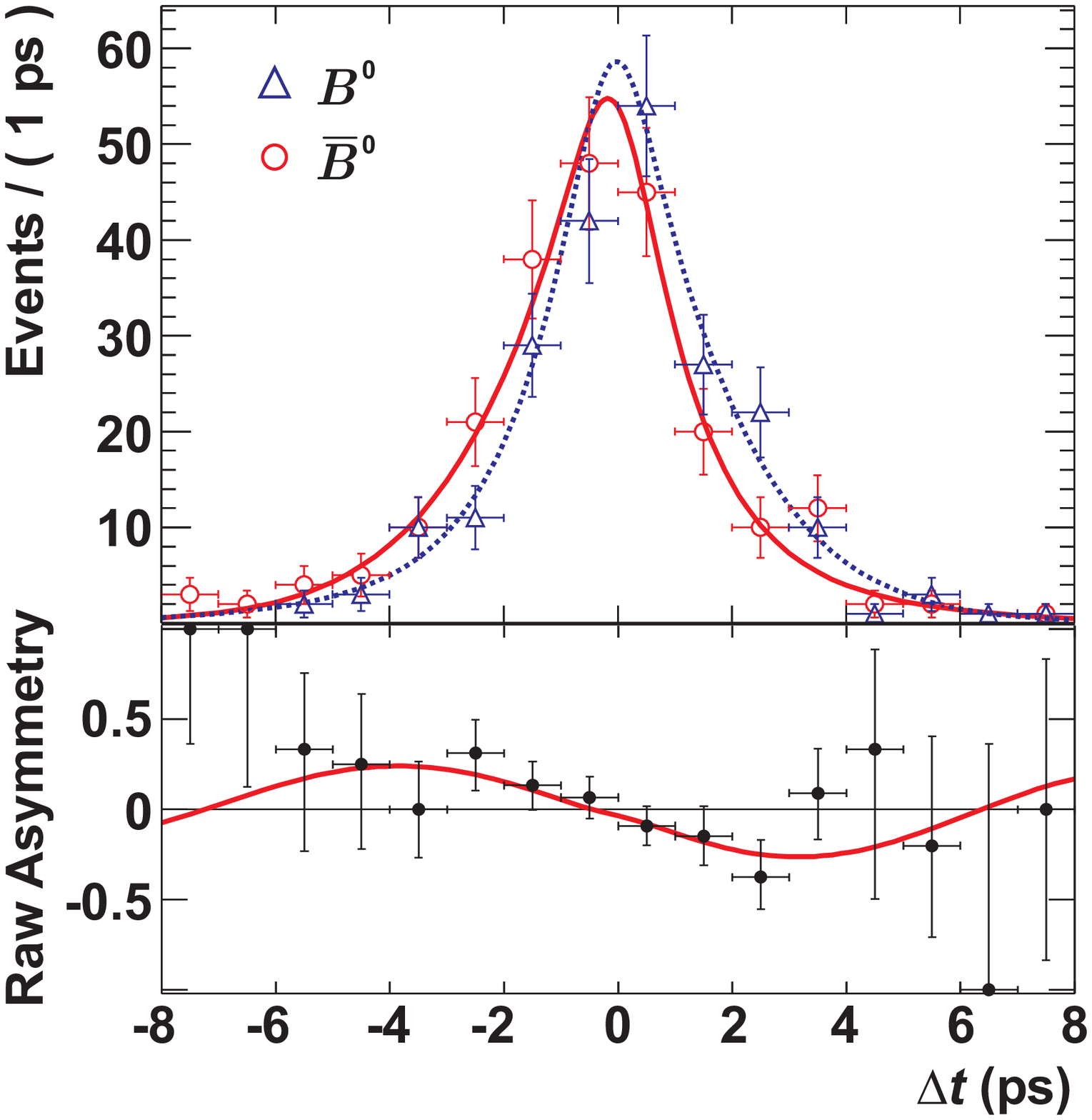}}\hspace{1.0in}
  \subfigure[~\Bztodd]{\includegraphics[width=0.75\columnwidth]{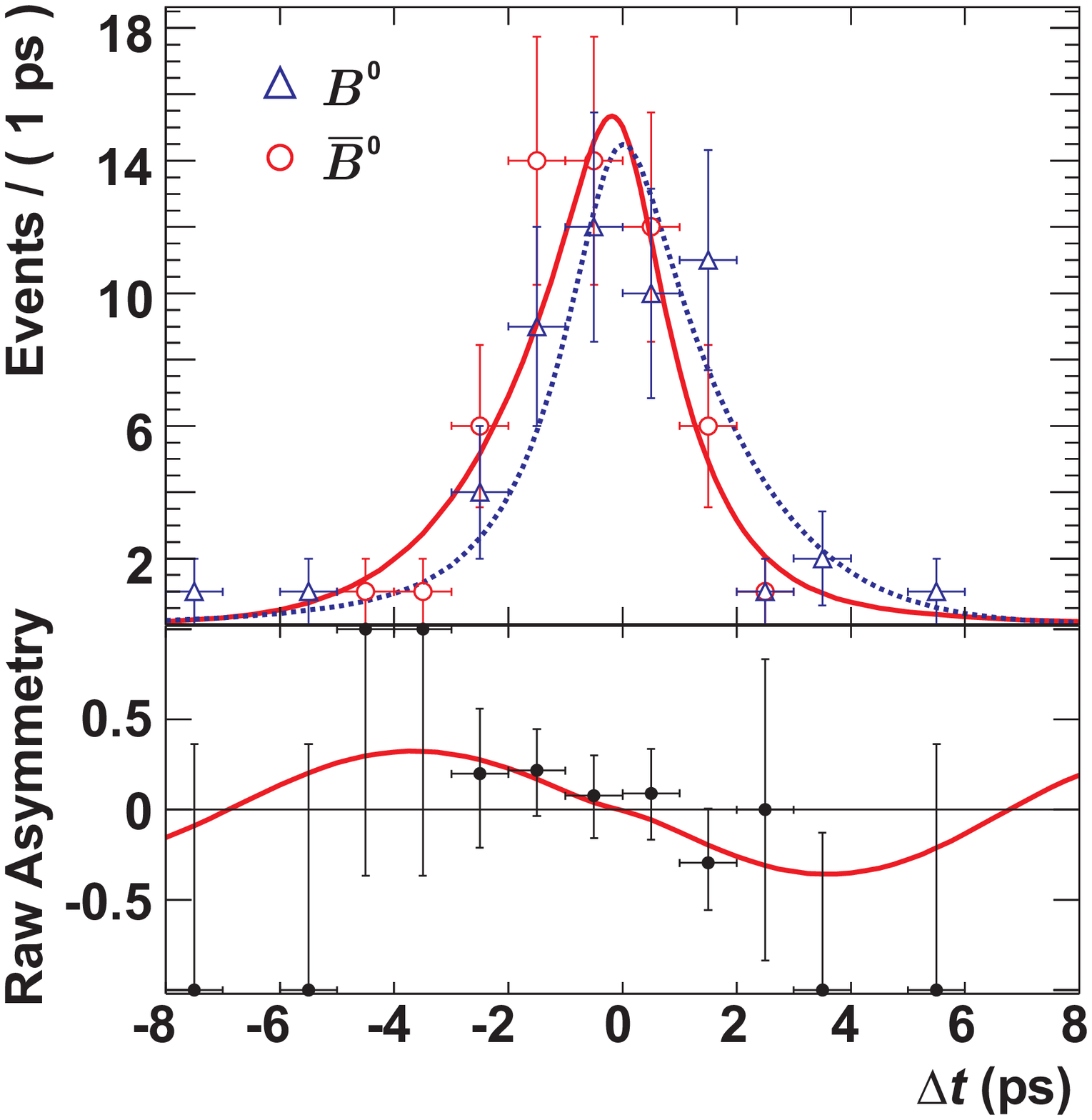}}
  \subfigure[~\Bztodstd]{\includegraphics[width=0.75\columnwidth]{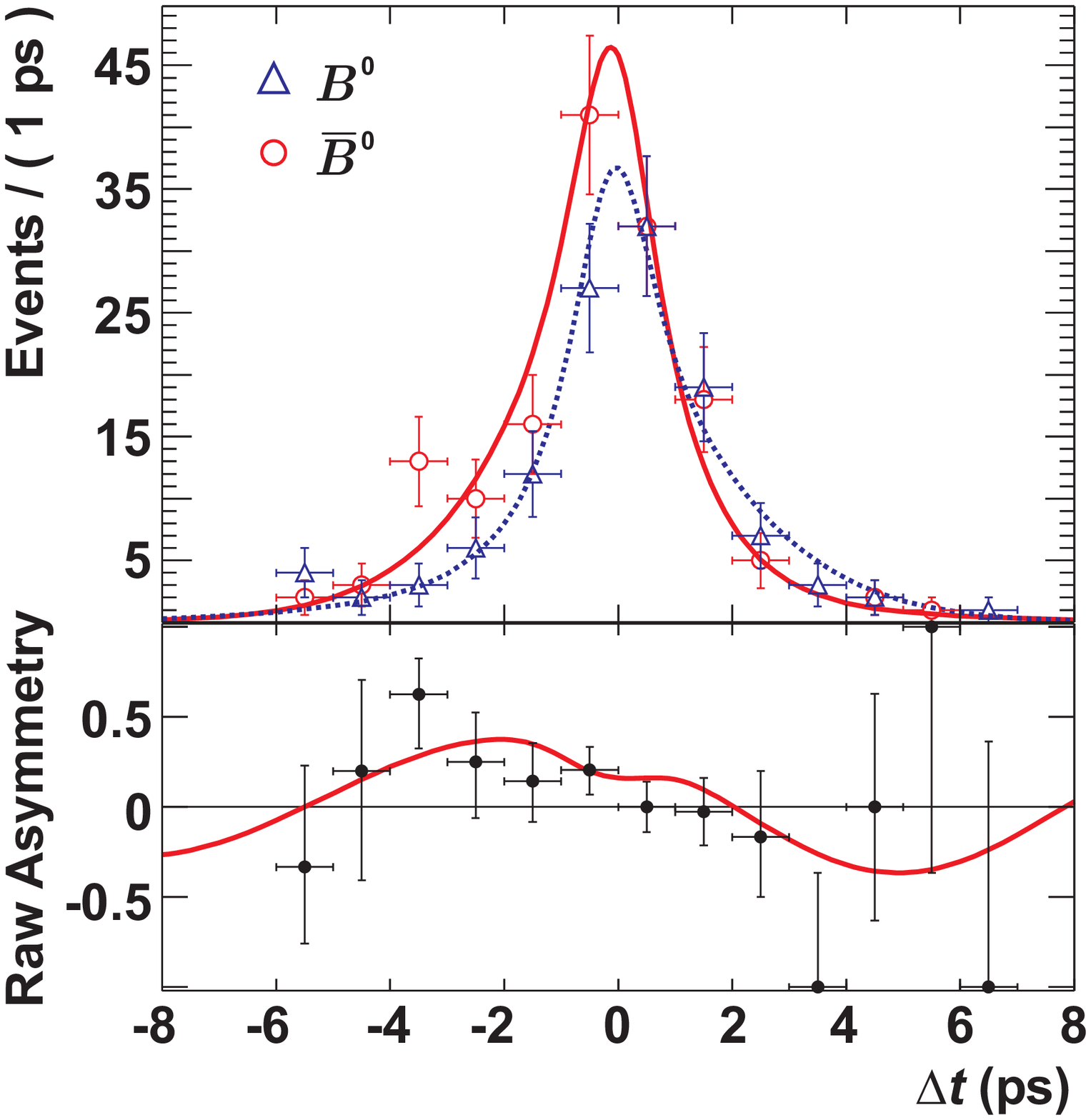}}\hspace{1.0in}
  \subfigure[~\Bztoddst]{\includegraphics[width=0.75\columnwidth]{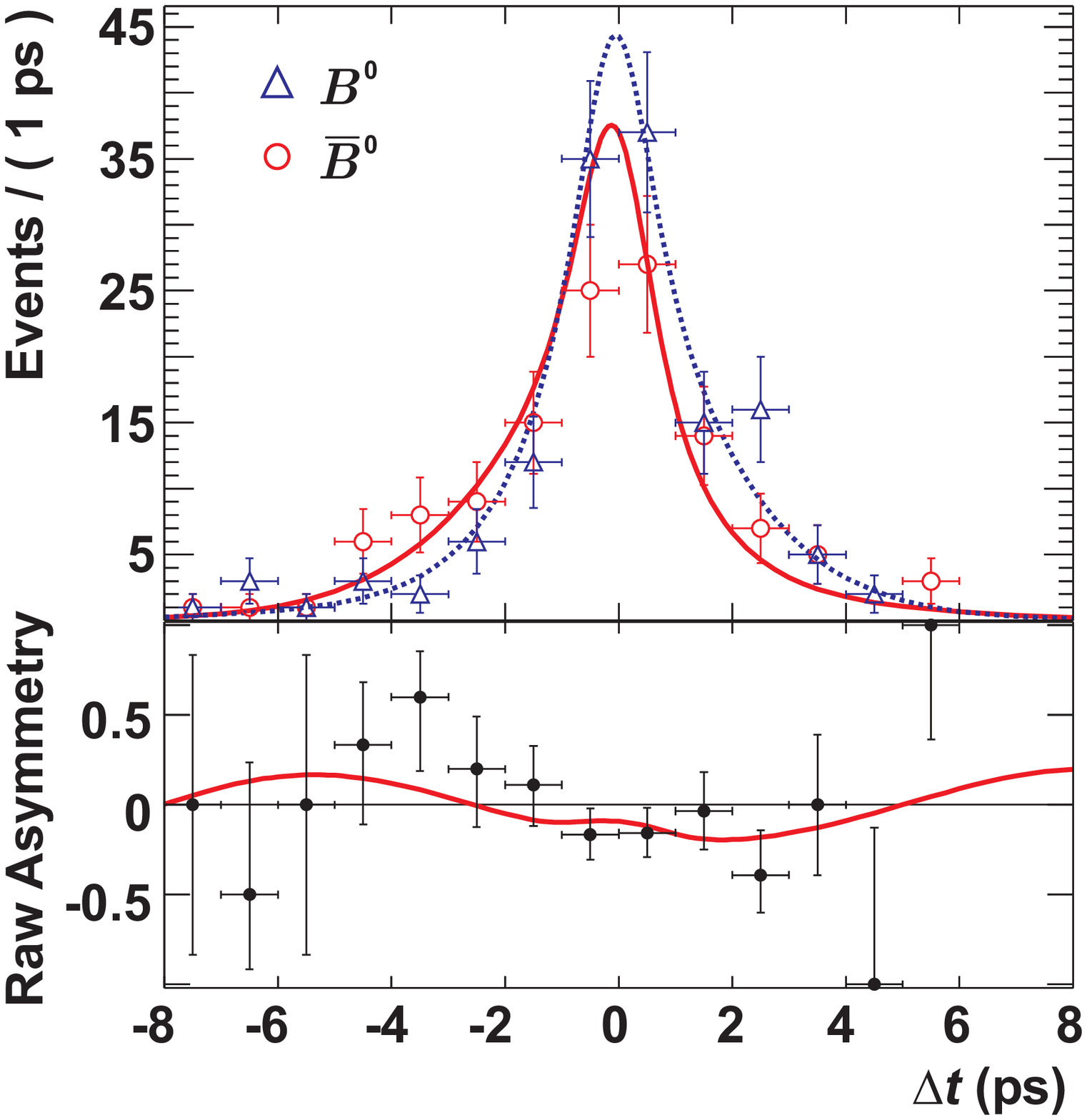}}
  \caption{\label{fig:dtasym}Projections onto \deltat of the fit
    result and the data in the region $\mes > 5.27\gevcc$ for the
    three highest purity tagging categories.  The triangular points
    and the dashed lines are for \Bz tagged events, and the circular
    points and solid lines are for \Bzb tagged events.}
\end{figure*}

We determine the \CP asymmetry parameters in unbinned ML fits to the
\mes, \deltat, and in the case of \Bztodstdst, \costhtr distributions.
The \deltat signal distributions are given in
Eqs.~\ref{eq:tddecayrate} and~\ref{eq:cprt} convolved with the
experimental resolution.  The \deltat background distribution has both
zero and nonzero lifetime components which are convolved with the
experimental resolution.  The lifetime component is allowed to have
effective \CP parameters and lifetime, which are determined in the
fits.  The angular measurement resolution, determined for the \CP-odd
fraction measurement, is convolved with the signal angular
distribution.  The efficiency moments are not modeled but rather
absorbed into an effective \Rt, which is determined in the fit.  This
procedure simplifies the \costhtr distribution and does not introduce
a bias.  The peaking background for the $\Bz\to D^{(*)\pm}D^\mp$
channels shares the \deltat background distributions with the
combinatorial background because it originates from similar sources.
The \Bztodstdst peaking background has only a lifetime component,
since it originates from a specific $\Bp$ decay.  Untagged events are
also included in the fits to constrain the \mes and \costhtr shapes
but do not contribute to the determination of the \CP parameters.  We
also allow the signal yield, the \mes background shape, and the
\costhtr background shape to vary in the fits.  Again we use
high-statistics MC samples divided into data-sized subsets to validate
the fitting procedure and find no significant bias.

The statistical uncertainties of the \CP measurements below are
consistent with the expected uncertainties obtained from MC studies
that include the signal and background yields observed in data.  The
statistical uncertainty for the $\Bz\to\Dstapm\Dmp$ channels is
essentially unchanged or even slightly worse than our previous
measurement~\cite{Aubert:2007pa}.  We interpret this as a downward
fluctuation in the statistical uncertainty of the previous
measurement.  Using MC data, we estimate the probability of observing
such a fluctuation at about 20\%.  For each measurement that follows,
the first uncertainty is statistical and the second is systematic.

From the fit to the \Bztodstdst data, we find
\begin{align}
  S_+ &= -0.76 \pm 0.16 \pm 0.04 \nonumber \\
  C_+ &= +0.00 \pm 0.12 \pm 0.02 \nonumber \\
  S_\perp &= -1.80 \pm 0.70 \pm 0.16 \nonumber \\
  C_\perp &= +0.41 \pm 0.49 \pm 0.08\,,
  \label{eq:cptr}
\end{align}
with an effective $\Rt = 0.155\pm 0.030$.  If we perform the fit with
the additional constraints that $S_+ = S_\perp = S_{\Dstarp\Dstarm}$
and $C_+ = C_\perp = C_{\Dstarp\Dstarm}$, we obtain
\begin{align}
  S_{\Dstarp\Dstarm} &= -0.70 \pm 0.16 \pm 0.03 \nonumber \\
  C_{\Dstarp\Dstarm} &= +0.05 \pm 0.09 \pm 0.02\,,
  \label{eq:cprt1S}
\end{align}
having an effective $\Rt = 0.171 \pm 0.028$.  Fitting the \Bztodd data
yields
\begin{align}
  S_{\Dp\Dm} &= -0.63 \pm 0.36 \pm 0.05 \nonumber \\
  C_{\Dp\Dm} &= -0.07 \pm 0.23 \pm 0.03\,,
  \label{eq:cpdd}
\end{align}
and fitting the $\Bz\to\Dstarpm\Dmp$ data yields
\begin{align}
  S_{\Dstarp\Dm} &= -0.62 \pm 0.21 \pm 0.03 \nonumber \\
  S_{\Dp\Dstarm} &= -0.73 \pm 0.23 \pm 0.05 \nonumber \\
  C_{\Dstarp\Dm} &= +0.08 \pm 0.17 \pm 0.04 \nonumber \\
  C_{\Dp\Dstarm} &= +0.00 \pm 0.17 \pm 0.03\,.
  \label{eq:cpdstd}
\end{align}
Projections of the fit results onto \deltat for events in the region
$\mes > 5.27 \gevcc$, and their flavor asymmetry, can be seen in
Fig.~\ref{fig:dtasym}.  To enhance the visibility of the signal in
these projections, we use three of the six tagging categories with the
highest purity, which account for 80\% of the total effective tagging
power $Q$.  The correlations among the \CP parameters are given in the
appendix.

\begin{table*}
  \caption{\label{tab:cprtsyst}Systematic uncertainties on the \Bztodstdst \CP 
    parameters.}
  \begin{ruledtabular}
    \begin{tabular}{lcccccc}
      & $S_+$ & $S_\perp$ & $C_+$ & $C_\perp$ & $S_{\Dstarp\Dstarm}$ & $C_{\Dstarp\Dstarm}$  \\
      \hline
      Tagging and \deltat resolution & 0.022 & 0.031 & 0.010 & 0.017 & 0.021 & 0.009 \\
      Peaking background & 0.012 & 0.079 & 0.002 & 0.019 & 0.012 & 0.003 \\
      Detector Alignment & 0.006 & 0.029 & 0.001 & 0.019 & 0.005 & 0.002 \\
      Doubly-Cabibbo suppressed decays & 0.002 & 0.002 & 0.014 & 0.014 & 0.002 & 0.014 \\
      Potential Fit Bias & 0.011 & 0.098 & 0.008 & 0.065 & 0.011 & 0.007 \\
      Angular PDF variations & 0.025 & 0.091 & 0.004 & 0.015 & 0.011 & 0.001 \\
      Other & 0.013 & 0.025 & 0.005 & 0.029 & 0.013 & 0.002 \\
      \hline
      Total & 0.040 & 0.163 & 0.020 & 0.080 & 0.032 & 0.018 \\
    \end{tabular}
  \end{ruledtabular}
\end{table*}

\begin{table*}
  \caption{\label{tab:cpsyst}Systematic uncertainties on the $\Bz\to
    D^{(*)\pm}\Dmp$ \CP parameters.}
  \begin{ruledtabular}
    \begin{tabular}{lcccccc}
      & $S_{\Dp\Dm}$ & $C_{\Dp\Dm}$ & $S_{\Dstarp\Dm}$ & $C_{\Dstarp\Dm}$ & $S_{\Dp\Dstarm}$ & $C_{\Dp\Dstarm}$ \\
      \hline
      Tagging and \deltat resolution & 0.031 & 0.011 & 0.027 & 0.012 & 0.029 & 0.011 \\
      \mes signal width & 0.034 & 0.020 & 0.013 & 0.018 & 0.028 & 0.012 \\
      Peaking background & 0.018 & 0.007 & 0.014 & 0.023 & 0.030 & 0.013 \\
      Detector Alignment & 0.002 & 0.001 & 0.004 & 0.002 & 0.002 & 0.001 \\
      Doubly-Cabibbo suppressed decays & 0.002 & 0.014 & 0.002 & 0.014 & 0.002 & 0.014 \\
      Potential Fit Bias & 0.007 & 0.005 & 0.008 & 0.006 & 0.008 & 0.006 \\
      Other & 0.006 & 0.002 & 0.002 & 0.006 & 0.004 & 0.003 \\
      \hline
      Total & 0.051 & 0.028 & 0.034 & 0.036 & 0.051 & 0.026 \\
    \end{tabular}
  \end{ruledtabular}
\end{table*}

We evaluate systematic uncertainties in the \CP asymmetries for each
mode by varying the fixed parameters for the mistag quantities and
\deltat resolution model within their uncertainties while accounting
for correlations among the parameters.  For the \Bztodd and
$\Bz\to\Dstarpm\Dmp$ modes, we change the fixed \mes signal width by
$\pm 0.2\mevcc$, an amount determined from a comparison of data and MC
event samples in modes with high purity, and take the difference in
fitted results as a systematic uncertainty.  Additionally, we vary the
fraction and shape of the peaking background component.  We also
include systematics for possible detector misalignment and the
presence of doubly-Ca\-bib\-bo suppressed decays of the $B_\text{tag}$
meson~\cite{Long:2003wq}.  We assign a systematic uncertainty equal to
the statistical uncertainty of the MC sample used to validate the fit.
Other sources of systematic uncertainty include: the \Bz meson
properties (\deltamd and $\tau_{\Bz}$), which we vary to $\pm 1
\sigma$ of their world averages, and uncertainty in the boost; the
corresponding changes in the \CP asymmetries are taken as the estimate
of the systematic uncertainties.  For the \Bztodstdst mode, we vary
the \costhtr resolution parameters and background shape in the manner
described for the evaluation of systematic uncertainties on \Rt and
take the effects on the \CP parameters as the associated systematic
uncertainty.  A summary of the systematic uncertainties for the \CP
parameters is given in Tables~\ref{tab:cprtsyst} and~\ref{tab:cpsyst}.
As before, the total systematic uncertainty is the sum in quadrature
of the individual contributions.

Because $\Bz\to\Dstarpm\Dmp$ decays are not \CP eigenstates, it is
illustrative to express the \CP asymmetry parameters $S$ and $C$ in a
slightly different parametrization~\cite{Aubert:2003wr}
\begin{align}
  S_{\Dstar D} &= \frac{1}{2}\left(S_{\Dstarp\Dm}+S_{\Dp\Dstarm}\right) \nonumber \\
  \Delta S_{\Dstar D} &= \frac{1}{2}\left(S_{\Dstarp\Dm}-S_{\Dp\Dstarm}\right) \nonumber \\
  C_{\Dstar D} &= \frac{1}{2}\left(C_{\Dstarp\Dm}+C_{\Dp\Dstarm}\right) \nonumber \\
  \Delta C_{\Dstar D} &=
  \frac{1}{2}\left(C_{\Dstarp\Dm}-C_{\Dp\Dstarm}\right)\,.
  \label{eq:avgcp}
\end{align}
The $S_{\Dstar D}$ and $C_{\Dstar D}$ parameters characterize
mixing-induced \CP violation related to the angle $\beta$ and
flavor-dependent direct \CP violation, respectively.  $\Delta
S_{\Dstar D}$ is insensitive to \CP violation but is related to the
strong phase difference $\delta$.  $\Delta C_{\Dstar D}$ describes the
asymmetry between the rates $\Gamma(\Bz\to\Dstarp\Dm) +
\Gamma(\Bzb\to\Dp\Dstarm)$ and $\Gamma(\Bz\to\Dp\Dstarm) +
\Gamma(\Bzb\to\Dstarp\Dm)$.  Using the results from
Eq.~\ref{eq:cpdstd} and taking into account correlations among the
variables, we find
\begin{align}
  S_{\Dstar D} &= -0.68 \pm 0.15 \pm 0.04 \nonumber \\
  \Delta S_{\Dstar D} &= +0.05 \pm 0.15 \pm 0.02 \nonumber \\
  C_{\Dstar D} &= +0.04 \pm 0.12 \pm 0.03 \nonumber \\
  \Delta C_{\Dstar D} &= +0.04 \pm 0.12 \pm 0.03\,.
  \label{eq:avgcpdstd}
\end{align}

From the signal yields $N_{\Dstarp\Dm}$ and $N_{\Dp\Dstarm}$
determined in the time-dependent fit described above, we also measure
the time-integrated \CP asymmetry in $\Bz\to\Dstarpm\Dmp$ decays,
defined as
\begin{equation}
  \mathcal{A} = \frac{N_{\Dstarp\Dm}-N_{\Dp\Dstarm}}{N_{\Dstarp\Dm}+N_{\Dp\Dstarm}}\,.
\end{equation}
We find
\begin{equation}
  \mathcal{A} = +0.008 \pm 0.048\stat \pm 0.013\syst\,,
  \label{eq:Acp}
\end{equation}
where the systematic uncertainty is dominated by track reconstruction
efficiency differences for positive and negative tracks (0.013).
There is also a small contribution from the \mes signal width, peaking
background, and MC statistics (0.002).

\section{Conclusion}
\label{sec:conc}

We have measured the \CP asymmetry parameters for \Bztodstadsta
decays, including the \CP-odd fraction in the \Bztodstdst channel,
using the final \babar\ data sample.  All of the $S$ parameters are
consistent with the value of \stwob measured in $b\to(\ccbar)s$
transitions~\cite{Aubert:2007hm} and with the expectation from the
Standard Model for small penguin contributions.  The $C$ parameters
are consistent with zero in all modes.  In particular, we see no
evidence of the large direct \CP violation reported by the Belle
Collaboration in the \Bztodd channel~\cite{Fratina:2007zk}.  This
measurement supersedes the previous \babar\
measurements~\cite{Aubert:2007pa,Aubert:2007rr} of \CP asymmetries in
these decays.

\begin{acknowledgments}
\label{sec:ack}
We are grateful for the extraordinary contributions of our \pep2\
colleagues in achieving the excellent luminosity and machine
conditions that have made this work possible.  The success of this
project also relies critically on the expertise and dedication of the
computing organizations that support \babar.  The collaborating
institutions wish to thank SLAC for its support and the kind
hospitality extended to them.  This work is supported by the U.S.
Department of Energy and National Science Foundation, the Natural
Sciences and Engineering Research Council (Canada), the Commissariat
\`a l'Energie Atomique and Institut National de Physique Nucl\'eaire
et de Physique des Particules (France), the Bundesministerium f\"ur
Bildung und Forschung and Deutsche Forschungsgemeinschaft (Germany),
the Istituto Nazionale di Fisica Nucleare (Italy), the Foundation for
Fundamental Research on Matter (The Netherlands), the Research Council
of Norway, the Ministry of Education and Science of the Russian
Federation, Ministerio de Educaci\'on y Ciencia (Spain), and the
Science and Technology Facilities Council (United Kingdom).
Individuals have received support from the Marie-Curie IEF program
(European Union) and the A. P. Sloan Foundation.
\end{acknowledgments}

\appendix*
\section{\boldmath Correlations among the \CP parameters}
\label{app:corr}

To allow detailed use of these results, we include the correlation
matrices for the \CP parameters.  Table~\ref{tab:corrdstdst} contains
correlations among the fit parameters in the \Bztodstdst channel with
separate \CP-even and \CP-odd asymmetries, and in the combined case,
the correlation between $S_{\Dstarp\Dstarm}$ and $C_{\Dstarp\Dstarm}$
is $0.8\%$ with correlations to the effective \Rt the same as the
\CP-even parameters.  Table~\ref{tab:corrdstd} contains the
correlations among the $\Bz\to\Dstarpm\Dmp$ asymmetries.  The
correlation of the time-integrated \CP asymmetry $\mathcal{A}$ with
any of the \CP parameters is less than $0.1\%$.  The correlation
between $S_{\Dp\Dm}$ and $C_{\Dp\Dm}$ is $-1.2\%$.

\begin{table}[tp]
\caption{\label{tab:corrdstdst}Correlations among the \CP parameters 
of the \Bztodstdst mode split by \CP-even and \CP-odd.}
\begin{ruledtabular}
\begin{tabular}{lccccc}
& $S_+$ & $C_+$ & $S_\perp$ & $C_\perp$ & \Rt \\
\hline
$S_+$ & 1 & $0.008$ & $0.376$ & $-0.036$ & $-0.083$ \\
$C_+$ & & 1 & $0.045$ & $-0.465$ & $\phantom{-}0.003$ \\
$S_\perp$ & & & 1 & $-0.224$ & $\phantom{-}0.471$ \\
$C_\perp$ & & & & 1 & $-0.151$ \\
\Rt & & & & & 1 \\
\end{tabular}
\end{ruledtabular}
\end{table}

\begin{table}[tp]
\caption{\label{tab:corrdstd}Correlations among the \CP parameters 
of the $\Bz\to\Dstarpm\Dmp$ mode.}
\begin{ruledtabular}
\begin{tabular}{lcccc}
& $S_{\Dstarp\Dm}$ & $C_{\Dstarp\Dm}$ & $S_{\Dp\Dstarm}$ & $C_{\Dp\Dstarm}$ \\
\hline
$S_{\Dstarp\Dm}$ & 1 & $-0.039$ & $\phantom{-}0.002$ & $\phantom{-}0.001$ \\
$C_{\Dstarp\Dm}$ & & 1 & $-0.001$ & $-0.001$ \\
$S_{\Dp\Dstarm}$ & & & 1 & $-0.009$ \\
$C_{\Dp\Dstarm}$ & & & & 1 \\
\end{tabular}
\end{ruledtabular}
\end{table}

\end{document}